# Low-temperature electron-spin relaxation in the crystalline and glassy states of solid ethanol

Marina Kveder, Dalibor Merunka, Milan Jokić, and Boris Rakvin

*Ruđer Bošković Institute, Bijenička 54, 10000 Zagreb, Croatia*

*X*-band electron paramagnetic resonance spectroscopy was used to study the spectral properties of a nitroxide spin probe in ethanol glass and crystalline ethanol, at 5–11.5 K. The different anisotropy of molecular packing in the two host matrices was evidenced by different rigid limit values for maximal hyperfine splitting in the signal of the spin probe. The significantly shorter phase memory time $T_m$ for the spin probe dissolved in crystalline ethanol, as compared to ethanol glass, was discussed in terms of contribution from spectral diffusion. The effect of low-frequency dynamics was manifested in the temperature dependence of $T_m$ and in the difference between the data measured at different spectral positions. This phenomenon was addressed within the framework of the slow-motional isotropic diffusion model [S. Lee and S. Z. Tang, Phys. Rev. B **31**, 1308 (1985)] predicting the spin probe dynamics within the millisecond range, at very low temperatures. The shorter spin-lattice relaxation time of the spin probe in ethanol glass was interpreted in terms of enhanced energy exchange between the spin system and the lattice in the glass matrix due to boson peak excitations.

## I. INTRODUCTION

The process of vitrification in glass-forming materials continues to represent a challenge for both experimental and theoretical physicists.[1] As the underlying dynamics in disordered and amorphous systems at low temperatures exhibits complexity at both the macroscopic and the microscopic levels, relaxation phenomena cannot be described within a self-consistent theory applicable to the whole frequency range. In this context, the phenomenon of boson peak (BP) excitations, describing an excess of lattice vibrational density of states, relative to the Debye regime, remains a challenge in professional debates, and different approaches are proposed to address the analysis of BP origin and related phenomena.[2–7] In electron paramagnetic resonance (EPR) spectroscopy, it is expected that BP excitations will influence the energy exchange of the electron spins with the lattice and, thus, contribute to the relaxation processes in the system.

Low-temperature effects have been exhaustively investigated in solid alcohols because they can be prepared in phases characterized by different types of disorder, which result from a controlled thermal history of the sample.[5–10] Exploiting the sensitivity of EPR spectroscopy to relaxation processes spanning the time scale from nanoseconds to milliseconds, here, we characterize crystalline ethanol and ethanol glass matrices by EPR in a low-temperature window (5–11.5 K). In this temperature range, the existence of BP excitations affects the physical properties of solid ethanol, such as its heat capacity.[3,5] The approach employed here is based on the analysis of the EPR spectral properties of a nitroxide paramagnetic spin probe incorporated in ethanol, a concept well established since the early work of Goldman *et al.*[11,12] However, to the best of authors' knowledge, EPR data obtained below 15 K are scarce, in this context,[13–16] even though they would be of particular significance for the validation of the theoretical assumptions.[17–19]

Continuing our previous EPR studies on solid ethanol phases, using nitroxide spin probes,[20] here, we employ a nitroxide exhibiting superior sensitivity toward the anisotropy of molecular packing,[20] and extend the study of electron-spin relaxation toward lower temperatures (5–11.5 K). Phase memory times $T_m$ and spin-lattice relaxation times $T_1^*$ were analyzed with regard to host matrix polymorphism. The temperature dependence of phase memory times provided experimental evidence that reporter groups embedded in glassy and crystalline environment exhibited different low-frequency dynamics. The theoretical analysis was based on the theory of slow orientational diffusional motion of EPR hyperfine centers in amorphous samples.[21] It was shown that the underlying dynamics of the system, characterized by correlation times in the range of milliseconds, existed even at cryogenic temperatures. The low-temperature spin-lattice relaxation times of the paramagnetic centers were different in different solid ethanol matrices. This phenomenon was discussed in terms of the impact of the density of vibrational excitations on the energy exchange between electron spins and the lattice.

## II. EXPERIMENT

Ethanol [anhydrous, minimum 99.8% [gas chromatography (GC)], pro analysis] was from Kemika, Zagreb. The liquid was doped with the nitroxide paramagnetic spin probe 2,2,6,6-tetramethyl-1-piperidine-1-oxyl (TEMPO) from Aldrich, at a concentration of 0.7 mM. The influence of trace amounts of water (less than 0.2% w/w) and of the spin probe (0.01% w/w) on solid ethanol polymorphism[8] can be neglected for the purpose of this study. The sample was deoxygenated in an argon atmosphere. Glassy and crystalline ethanol were prepared as described.[20]

EPR measurements were performed using a Bruker E-580 Fourier transform (FT)/cw *X*-band spectrometer equipped with an Oxford Instruments temperature unit (±0.1 K). For cw-EPR measurements at low temperatures, the acquisition parameters were carefully optimized as follows: microwave power of 0.0002 mW and modulation frequency of 10 kHz; the modulation phase was adjusted so as to minimize the contribution from the signal out of phase. The phase memory time $T_m$ was derived from two-pulse electron-spin-echo (ESE) decay[22] with a pulse separation time of 200 ns and a $\pi$ pulse duration of 88 ns. The spin echo dephasing curves deviated from single-exponential shape and were approximated by stretched exponential functions.[14,15] Measurements were performed at three spectral positions deduced from a field-swept ESE detected experiment[22] at 5 K. Spin-lattice relaxation times were determined by inversion recovery using an echo detection sequence[22] with the same pulse separation time and $\pi$ pulse duration as those in the $T_m$ measurements. As spin-lattice relaxation times at very low temperature increase above the limits accessible to current instrumentation, an "effective" spin-lattice relaxation time $T_1^*$ was used. As shown and discussed previously,[23] this was deduced from a biexponential fit wherein only the longer component was considered an approximation for $T_1^*$.

All calculations were performed with the MATHEMATICA 5.1 (Wolfram) software package.

## III. RESULTS

cw-EPR spectra were used to estimate maximal hyperfine splitting, $2A_{max}$,[20] in the temperature range (5–11.5 K) studied. TEMPO incorporated in crystalline ethanol showed a larger value, $2A_{max}=7.4$ mT, than TEMPO incorporated in ethanol glass, $2A_{max}=7.2$ mT. Since the anisotropy of molecular packing is smaller in ethanol glass than in crystalline ethanol, $2A_{max}$ should be smaller in the former than in the latter sample type. These $2A_{max}$ values did not appear to change within the chosen temperature interval and, thus, can be assigned to the apparent rigid limit values at the level of the conventional X-band cw-EPR. This interpretation does not imply that the nitroxide is immobilized, not even at these cryogenic temperatures, but rather that the underlying dynamics is expected to be slower than 10 $\mu$s and should be studied by pulsed EPR measurements.

### A. Temperature dependence of the phase memory relaxation time

The phase memory time $T_m$ of the spin probe TEMPO in different phases of solid ethanol was measured at 5–11.5 K. Three spectral positions (low, central, and high fields) for the excitations of the spin packets at a particular temperature were taken from the field-swept ESE experiment. The typical experimental two-pulse ESE decay curves are shown in Fig. 1 for the lowest measured temperature. The relaxation curves were fitted to the stretched exponential function

$$I(t) = I(0)e^{-(t/T_m)^x}, \quad (1)$$

wherein $I(t)$ is the intensity of the echo at time $t$. Equation (1) has frequently been employed in the evaluation of low-temperature EPR relaxation decay data, using values of the exponent $x$ greater than 1.[14,15] This is explained by the incomplete excitation of the spin system in the solid sample by the microwave pulses. As a consequence, an irreversible dephasing of the excited spin frequencies is induced and reflects on the time course of the decay of spin echo signals. In the physical models so far proposed for the evaluation of electron-spin-echo decay in rigid samples, the exponent $x$ has been assigned values ranging from 0.5 to 2.[15,24] The strong departure of ESE decay from the monoexponential behavior observed in the present study may, in part, result from slow molecular motion and/or various mechanisms of spectral diffusion[25,26] and from specific distributions of correlation times at the level of local relaxation events.[27] The inverse of the phase memory time $1/T_m$, derived according to Eq. (1), is presented in Fig. 2. A clear difference between the data for ethanol glass and crystalline ethanol can be observed, indicating longer $T_m$ for the former than for the latter type of the sample, throughout the whole temperature range studied. Within the same type of the solid ethanol matrix, the $1/T_m$ values at a particular temperature are different when measured at different spectral positions (Fig. 2). For both ethanol matrices, the TEMPO phase memory time increases as the temperature decreases.

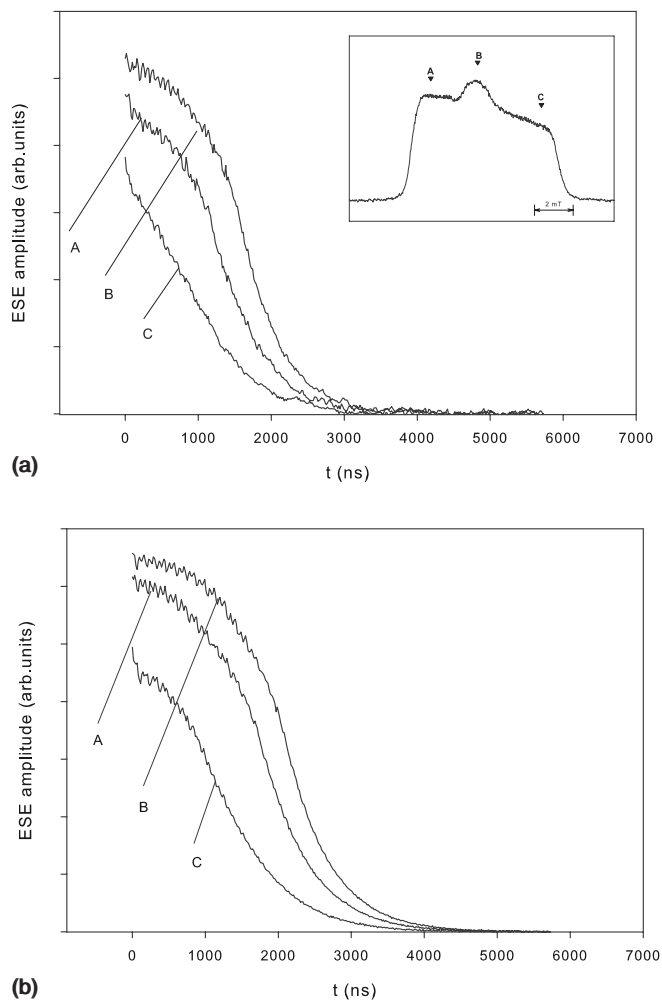

FIG. 1. ESE decay curves of TEMPO in (a) crystalline ethanol and (b) ethanol glass observed at 5 K. A field-swept ESE signal is shown in the inset to the figure to indicate low (A), central (B), and high (C) magnetic field positions at which the experimental data were acquired.

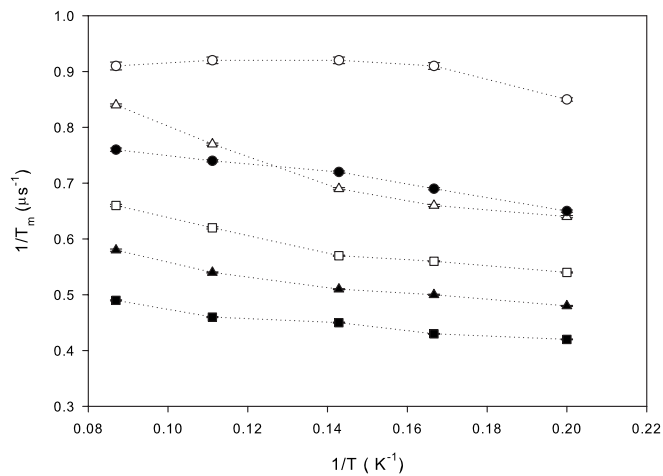

FIG. 2. Temperature dependence of $1/T_m$ data of TEMPO incorporated in ethanol glass (solid symbols) and crystalline ethanol (plain symbols) derived by fitting the experimental data to Eq. (1). The measurements were performed at magnetic field positions A (triangles), B (squares), and C (circles). The dotted lines serve to guide the eyes.

### B. Temperature dependence of the spin-lattice relaxation time

The spin-lattice relaxation rate of TEMPO incorporated in different solid ethanol matrices was measured at the central magnetic field position in the temperature window from 5 to 11.5 K, as shown in Fig. 3. $T_1^*$ in ethanol glass was significantly shorter than $T_1^*$ in crystalline ethanol for all temperatures studied. This is consistent with previous measurements at higher temperatures.[20] No hysteresis was observed for $T_1^*$, irrespectively of whether the target sample temperature was reached by heating or by cooling. Approaching 5 K, the temperature dependence of the spin-lattice relaxation times in the two solid ethanol phases almost leveled off.

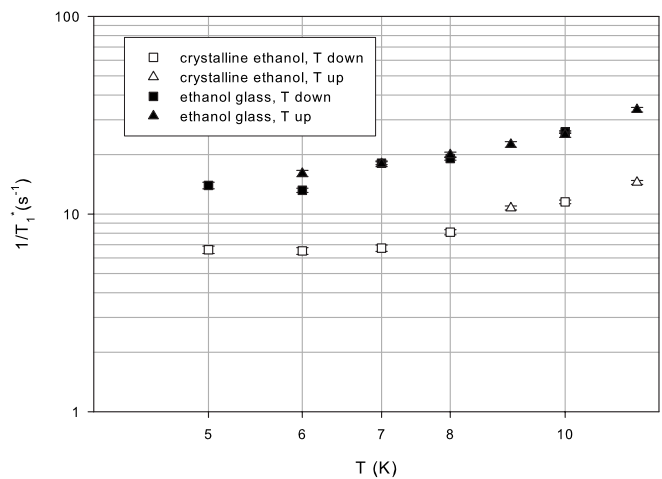

FIG. 3. Temperature dependence of $1/T_1^*$ for TEMPO incorporated in ethanol glass (solid symbols) and crystalline ethanol (plain symbols) measured at the central magnetic field position. The temperature range was investigated by both decreasing (square symbols) and increasing (triangle symbols) the temperature. The experimental error bars are within the size of the symbols.

No attempt was made to simulate the temperature dependence of $T_1^*$ in terms of the expected low-temperature linear behavior[23] or tunneling level states,[28] due to the effective character of the measured spin-lattice relaxation times.

## IV. DISCUSSION

Here, we show that pulsed EPR provides information on the underlying dynamics in solid ethanol matrices at low temperatures (5–11.5 K). Phase memory and spin-lattice relaxation times of paramagnetic centers embedded in ethanol glass and crystalline ethanol were investigated. Solid ethanol exists in a number of interesting polymorphs, such as orientationally disordered crystal with glassy properties, sometimes named *glassy crystal*, which shows low-temperature specific heat values very similar to those of ethanol glass.[29] In this study, only two extreme cases, the glass and the crystalline matrix, were considered as they could be unambiguously and reproducibly identified from the thermal history of the spectra.

In the studied temperature range (5–11.5 K), the apparent rigid limit hyperfine splitting values, derived from X-band cw-EPR experiments, indicated that the underlying dynamics, if contributing, is expected to be slower than 10 μs and, thus, should be studied by pulsed EPR techniques.

### A. Low-temperature dynamics of paramagnetic centers in solid ethanol matrices

The question of motional dynamics of paramagnetic centers was addressed by analyzing the temperature dependence of phase memory time data. In pulsed EPR experiments, a narrow "hole" is burned into the spectral line, which is otherwise inhomogeneously broadened, in such a way that $1/T_m$ can be approximated as

$$\frac{1}{T_m(M,\theta,\varphi)} = \frac{1}{T_2^{sd}} + \frac{1}{T_2^*(M,\theta,\varphi)}, \quad (2)$$

wherein $M$ denotes the nuclear spin quantum number and $\theta, \varphi$ are the angular spherical coordinates of the external magnetic field with respect to a coordinate system defined by the principal axes of the nitroxide moiety. The component $1/T_2^{sd}$ represents the contribution from spectral diffusion broadening and can be assumed to be equal for different spectral positions at a defined temperature.[30] The term $1/T_2^*(M,\theta,\varphi)$ describes the contribution of dynamics that modulates electron-nuclear dipolar interactions since other contributions, such as electron-electron dipolar interaction, can be neglected in the context of this study due to the dilute spin probe concentration applied. The importance of the $1/T_2^{sd}$ and $1/T_2^*(M,\theta,\varphi)$ terms is discussed as follows.

#### 1. Spectral diffusion in ethanol glass and crystalline ethanol

The experimental evidence of consistently longer $T_m$ values of the reporter groups in ethanol glass than crystalline ethanol can be explained by the effect of spectral diffusion, represented in Eq. (2) by the term $1/T_2^{sd}$. This phenomenon is a consequence of different hyperfine interactions of the unpaired electron in TEMPO with the nitrogen nuclei and different spatial averaging of dipolar interactions with the surrounding intra- and/or intermolecular protons due to different molecular packing densities in crystalline versus glassy host matrix. These protons serve as additional sensors of the spin probe accommodation in different physical states of the solvent cage. As the spin probe concentration was the same in both solid ethanol samples, and there was no experimental evidence of paramagnetic clustering or aggregation, the shorter $T_m$ values of TEMPO in crystalline ethanol most likely imply higher local concentrations of protons in the host matrix. Such a conclusion would be in accordance with the observation that crystalline ethanol is ~1.15 times more dense than ethanol glass.[5] Further supportive evidence includes the results of transverse field measurements in muon spin relaxation spectroscopy[9] and nuclear magnetic resonance (NMR) data,[31] which show larger second moments of the proton NMR line in crystalline ethanol than in ethanol glass. Since EPR lines are almost ten times narrower than the respective NMR lines,[31] the spin probe approach is more sensitive than NMR in the investigation of local differences in the lattice of two host matrices, especially at cryogenic temperatures.

### 2. Dynamic effects and the slow orientational diffusion model

The pronounced decrease in $1/T_m$ with decreasing temperature and the differences in the $1/T_m$ values obtained in measurements at different magnetic field strengths are consistent with the dynamic phenomena elaborated in the simple theoretical model described in Refs. 21 and 30. This approach was chosen due to acceptable underlying physical assumptions describing the dynamics of paramagnetic centers at low temperatures in terms of slow isotropic orientational diffusion motion in a restricted angular interval during EPR spin-spin relaxation time. The formalism was shown to be compatible with other models in the literature capable of reproducing EPR spectra in the regime of slow-motional dynamics.[32–34]

To extract the contribution of the dynamics that modulates the electron-nuclear dipolar interaction from the experimental data for phase memory time, the effect of spectral diffusion needs to be eliminated. This can be achieved by taking the difference between the high and low field hyperfine data while assuming that the influence of spectral diffusion is equal for different spectral positions at a particular temperature.[30] This reasoning, based on Eq. (2), can be formulated as

$$\frac{1}{T_m(M=-1)} - \frac{1}{T_m(M=+1)} = \Delta\left[\frac{1}{T_2^*}\right], \quad (3)$$

and the experimental data presented accordingly (Fig. 4). In line with the applied model, the excitations of the spin packets at spectral positions denoted as A and C in Fig. 1 were considered close to the contributions from the parallel-edge orientation ($\theta=0$) of nitroxide radicals for the $M=+1$ and $M=-1$ nitrogen hyperfine components, respectively. Referring to the theoretical expression for the spin-spin relaxation rate, $1/T_2$,[30] the following approximation for Eq. (3) can be proposed:

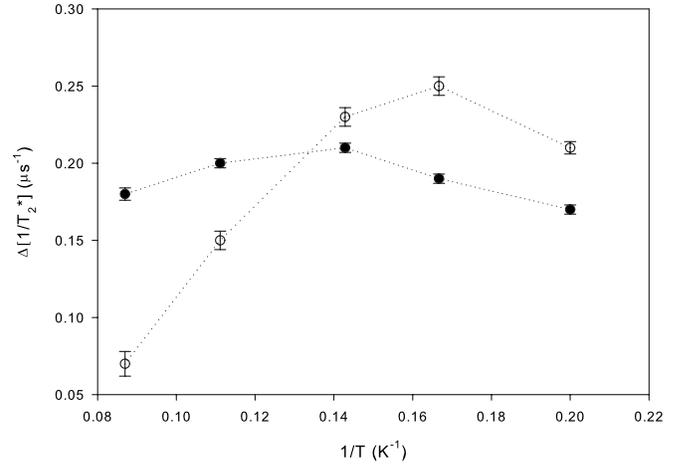

FIG. 4. Analysis of spin-spin relaxation rate data in the context of the applied model (Refs. 21 and 30). Temperature dependence of the parameter $\Delta[1/T_2^*]$ defined in Eq. (3). Solid (plain) symbols denote experimental data for TEMPO incorporated in ethanol glass (crystalline ethanol). The dotted lines serve to guide the eyes.

$$\Delta\left[\frac{1}{T_2^*}\right] \approx \Delta\left[\frac{1}{T_2}\right] = 2^{1/4}(5/36)^{1/2}(|\Delta_{M=-1}|^{1/2} - |\Delta_{M=+1}|^{1/2})\tau^{-1/2},$$

$$\Delta_M = 2\pi\hbar\nu_0(g_\perp^2 - g_\parallel^2)/g_\parallel + M(A_\perp^2 - A_\parallel^2)/A_\parallel, \quad (4)$$

wherein the hyperfine tensors $g$ and $A$ are assumed to be axially symmetric, $\tau$ is the correlation time of the dynamic process, and $\nu_0$ is the X-band EPR frequency. This approach predicts a decrease in $\Delta[1/T_2]$ with increasing correlation time, due to the lowering of the temperature. The experimental data presented in Fig. 4 follow the prediction of the applied theoretical model, at temperatures not exceeding 7 (6) K for ethanol glass (crystalline ethanol). Up to these temperatures, the motional correlation time of TEMPO in both solid ethanol matrices can be estimated from Eq. (4) resulting in values on a millisecond scale, corresponding to activation energies around 10 K. (Typical nitroxide tensor values were used in the calculation,[30] except for $A_\parallel$ which was approximated from the rigid limit hyperfine value.)

The reason why, at a specific temperature, the experimental data shown in Fig. 4 start deviating Eq. (4) appears to be beyond the scope of the applied model. A somewhat more general model, assuming EPR centers with principal $z$ axes oriented at arbitrary angles $\theta$, predicts the same phenomenon of decreasing $\Delta[1/T_2(\theta)]$ with increasing correlation time (data not shown). A more complex model for the calculation of $1/T_2$ should thus be searched which would permit the implementation of anisotropic $\tau$ and/or temperature dependence of the effective values of the tensors $A(\theta,\varphi)$ and $g(\theta,\varphi)$. In addition, the possibility for different distributions of correlation times depending on the architecture of the solid ethanol cage, as previously suggested in the description of spin probe motion in amorphous polymers, cannot be excluded.[35] In fact, the experimentally observed difference between paramagnetic relaxation in the crystalline versus the glassy host matrix strongly points to complexities in the dynamics of differently packed molecular environments.

### B. Energy exchange between the spins and the lattice

The experimental data for the spin-lattice relaxation time indicated longer $T_1^*$ for TEMPO incorporated in crystalline ethanol than in ethanol glass. This result is in accordance with NMR data[31] focused on the collective dynamics of the host matrix but necessarily obtained at much higher temperatures than the results presented here because the resonance of ethanol saturates at 80 K and below. The observed difference between the EPR data for ethanol glass and crystalline ethanol is much larger (a ratio close to 2 for the $1/T_1^*$ values), in the temperature range investigated here (5–11.5 K), than for previous data measured at higher temperatures (20–80 K).[20] The different $T_1^*$ of TEMPO in solid ethanol matrices could be explained by the reported BP low-frequency vibrations in ethanol glass, which persist until the vibrational lattice is destroyed around the transition to the liquid state.[36] Due to the higher density of low vibrational states in the glassy lattice, the energy exchange between the spins and the lattice is intensified, causing shorter $T_1^*$ in ethanol glass. This reasoning is corroborated by the results of other experimental approaches aimed at elucidating the differences in the physical properties of glassy and crystalline ethanol.[18] A direct observation of BP by EPR is beyond the scope of this study and will be the subject of future research.

## V. CONCLUSION

Two different solid ethanol phases were investigated by EPR, at low temperatures (5–11.5 K). The data obtained document the influence of structural disorder on the apparent rigid limit values. $T_m$ measurements permitted estimates of motional correlation times on a millisecond scale and of the, relatively small, energy of activation for paramagnetic group dynamics, in both solid ethanol states. These results confirm that pulsed EPR experiments are suitable tools permitting access to dynamic phenomena in glass and/or crystalline samples, at very low temperatures (below 10 K).

The smaller $T_1^*$ values for ethanol glass may be rationalized by the assumption that more pronounced boson peak excitations, due to the higher density of vibrational states, enhance energy exchange between the spins and the lattice, thus leading to a faster spin-lattice relaxation.

To extend this study to atomic level resolution, isotope effects will be investigated in future experiments to gain quantitative insight into spectral diffusion. For possible direct observation of BP influence on the spin-lattice relaxation, higher-frequency EPR techniques should be considered.

### ACKNOWLEDGMENTS

The authors would like to thank Janja Makarević for preparing the ethanol samples. This work was supported by the Croatian Ministry of Science, Education and Sports, Grants Nos. 098-0982915-2939 and 098-0982904-2912.